\documentstyle[12pt,aasms4]{article}

\begin{document}

\title{Hydrodynamics of GRB Afterglow}
\author{Re'em Sari}
\affil{Racah Institute, Hebrew University, Jerusalem 91904, Israel \\
and\\
Institute for Advanced Study, Princeton, NJ 08540 
}
\authoremail{sari@shemesh.fiz.huji.ac.il}

\begin{abstract}
The detection of delayed emission at X-ray optical and radio
wavelengths, ``after-glow'', suggests that the relativistic shell
which emitted the initial GRB due to internal shocks decelerates on
encountering an external medium, giving rise to the after-glow. We
explore the interaction of a relativistic shell with a uniform inter
stellar medium (ISM), up to the non relativistic stage. We demonstrate
the importance of several effects that were previously ignored, and
must be included in a detailed radiation analysis. At a very early
stage (few seconds), the observed bolometric luminosity increases as
$t^2$. On longer time scales (more than $\sim 10\sec$), the luminosity
drops as $t^{-1}$. If the main burst is long enough, an intermediate
stage of constant luminosity will form. In this case, the after-glow
overlaps the main burst, otherwise there is a time separation between
the two. On the long time scale, the flow decelerate in a self similar
way, reaching non relativistic velocities after $\sim30$
days. Explicit expressions for the radial profiles of this self
similar deceleration are given. Due to the deceleration and the
accumulation of ISM material, the relation between the observed time,
the shock radius, and its Lorentz factor, is given by
$t=R/16\gamma^2c$ which is a factor of $8$ different from the usual
expression. The majority of particles are those of the original ejecta
(and not the ISM) up to about $\sim 900$s. These particles reach
sub-relativistic velocities on a time scale of $\sim 2$hours, well
before the flow becomes sub-relativistic. Therefore the ejecta
particles are probably unimportant for most of the after-glow
radiation. We show that even though only a small fraction of the
energy is given to the electrons, most of the energy can be radiated over 
time. If this fraction is greater than $\sim 10\%$ radiation losses will 
significantly influence the hydrodynamical evolution at early times 
(less than $1$day).
\end{abstract}


\section{Introduction}

The isotropy of GRBs angular distribution combined with the
non-homogeneous distribution suggests that GRBs originate from
cosmological distances and therefor radiate energies of order of
$10^{51}$erg. Considerations of optical depth then show that the
bursts are produced by the dissipation of kinetic energy of highly
relativistic shells with Lorentz factor $\eta>100$ (see
\markcite{Piran}Piran 1996 for review). This dissipation can be either
due to internal shocks or due to the surrounding
ISM. \markcite{SP97}Sari \& Piran (1997) have shown that deceleration
on the ISM could not give rise to the variability observed in the
bursts (unless the process is very inefficient and involves much more
than $10^{51}$erg), while internal shocks could produce efficiently
the observed fluctuations (\markcite{KPS}Kobayashi, Piran and Sari
1997), if the ``inner engine'' has considerable fluctuations. It is
therefore likely that the main GRB is due to internal shocks.

These cosmological models predict that after the main GRB event the
ejecta decelerate due to interaction with the ISM, emitting radiation
at longer and longer wavelengths
(\markcite{PaczynskiRhoads}Paczy\'nski \& Rhoads 1993,
\markcite{Katz}Katz 1994, \markcite{MeszarosRees97}M\'esz\'aros \&
Rees 1997). This emission has been detected recently for several GRBs
due to an accurate determination of their position. The quantitative
agreement between the deceleration models and the measurements is good
(\markcite{Waxman97a}Waxman 1997a, Wijers, Rees \& M\'esz\'aros 1997,
\markcite{Waxman97b}Waxman 1997b). Since the quality of data for the
after-glows is higher than for the burst itself, more quantitative
results are needed.

In this letter we explore the hydrodynamics of the deceleration of a
relativistic fireball on a uniform ISM. We discuss the relation
between the rise time of the after-glow and the time of the main
burst. We use the analytic solution found by
\markcite{BlandfordMcKee}Blandford and McKee (1976) to describe the
swept up ISM and derive an expression for the position of the original
ejecta and its Lorentz factor. We show that although only a small
fraction of the internal energy is given to the radiating electrons, a
considerable amount of the energy can be radiated over the deceleration
period.

\section{The Rise of The After-Glow}
The problem of deceleration of a relativistic shell onto the ISM is
determined by four parameters: the initial shell's Lorentz factor
$\eta$, the energy of the shell $E=E_{52}10^{52}$erg, the width of the
shell (observer frame) $\Delta$ and the ISM density $n= n_1{\rm
cm}^{-3}$. The basic details of the interaction between the shell and
ISM where given in \markcite{SP95}Sari \& Piran 1995 and we briefly
review the main ideas. The treatment in this section is approximate
and correction factors of order unity may need to be included in a
more precise treatment.

When the shell encounters the ISM, two shocks are formed: a forward
shock accelerating the ISM and a reverse shock decelerating the
shell. The forward shock is always highly relativistic since the
initial Lorentz factor $\eta\gg1$. Let $f$ be the density ratio
between the shell and the ISM given by
\begin{equation}
f=E/16 \pi \eta^2 \Delta n m_p c^4 \gamma^4 t^2.
\end{equation}
where we used $R=2c\gamma^2 t$ for the radius of the shell, t is the
observed time. The reverse shock is relativistic if $f<\eta^2$
reducing the Lorentz factor of the shell to
$\gamma=\eta^{1/2}f^{1/4}/\sqrt{2}$ and Newtonian if $f>\eta^2$ making
only negligible change to the shell's Lorentz factor i.e.,
$\gamma=\eta$. At early stages, $f>>\eta^2$ so the reverse shock is
Newtonian and the shell's Lorentz factor equals its initial value
$\gamma=\eta$. However due to the increase in the area of the shell,
it produces internal energy in an increasing rate of
\begin{equation}
\label{power}
L=32 \pi c^5 n m_p \gamma^8 t^2
 =2.5\times 10^{50} \gamma_{300}^8 n_1 t_s^2 {\rm\ \frac{erg}{s}},
\end{equation}
where we use $t_s$ for the time in seconds. Assuming that the cooling
is fast (\markcite{SNP}Sari, Narayan \& Piran 1996), the observed
bolometric luminosity is proportional to the internal energy increase
rate and is therefore also given by Eq. (\ref{power}).  This behavior
will continue until either the shell has given the ISM an energy
comparable to its initial energy at
\begin{equation}
\label{tenewtonian}
t_E=\left( { 3E \over 32 \pi c^5 n m_p \eta^8} \right)^{1/3}
   =5\ E_{52}^{1/3}\eta_{300}^{-8/3}n_1^{-8/3}{\rm\ s},
\end{equation}
or until the reverse shock is no longer Newtonian, i.e. $f=\eta^2$ at
\begin{equation}
t_N=\left( {E \over 16 \pi \Delta n m_p c^4 \eta^8} \right)^{1/2}
 =2\ E_{52}^{1/2} \left( \frac \Delta {6\times 10^{11}{\rm cm}} \right)^{-1/2} \eta_{300}^4 n_1^{-1/2} {\rm\ s},
\end{equation}
whichever comes first. As in \markcite{SP95}Sari \& Piran (1995) we define the ratio
between the two expressions as
\begin{equation}
\xi\equiv{t_N \over t_E}=
 \left(   {E \over 36\pi n m_p c^2 \Delta^3 \eta^8}  \right)^{1/6}
=0.4\ E_{52}^{1/6} \left( \frac \Delta { 6\times 10^{11}{\rm cm} } \right)^{-1/2}
 \eta_{300}^{-4/3} n_1^{-1/6}.
\end{equation}
For $\xi>1$, the energy is dissipated to internal energy before the
Lorentz factor of the shell is reduced considerably. If $\xi<1$, then
the reverse shock turns relativistic before the kinetic energy of the
shell was emitted. In this case the Lorentz factor decreases with time
according to
\begin{equation}
\gamma(t)=\eta^{1/2}f(t)^{1/4}/\sqrt{2}
=300\ E_{52}^{1/8}\left( \frac \Delta {6\times 10^{11}{\rm cm}} \right)^{-1/8} n_1^{-1/8} t_s^{-1/4}.
\end{equation}
Note that at this stage the Lorentz factor $\gamma$ is independent of
its initial value $\eta$. Substituting this in Eq. (\ref{power}) we
get that the luminosity is constant in time and given by:
\begin{equation}
\label{powerrelativistic}
L={E \over 2\Delta/c}.
\end{equation}
This stage will continue until the shell has given the shocked ISM energy
comparable with its own at
\begin{equation}
\label{terelativistic}
t_E={2\Delta/c}.
\end{equation}
At this time the shell has decelerated to Lorentz factor of
\begin{equation}
\gamma=\left(\frac E {256 \pi \Delta^3 n m_p c^2} \right)^{1/8}=
120\ E_{52}^{1/8}\left( \frac \Delta {6\times 10^{11}{\rm cm}} \right)^{-3/8}n_1^{-1/8} ,
\end{equation}
independent of the initial Lorentz factor $\eta$. After the time
$t_E$, given by Eq. (\ref{tenewtonian}) or
\ref{terelativistic} the flow will be described by a self similar solution
as the ISM energy is now constant and comparable to the initial energy
of the shell $E$. As we shall show in the next section, from this time
$\gamma \propto R^{-3/2} \propto t^{-3/8}$ and therefore the observed
luminosity decreases as $t^{-1}$. This behavior is illustrated in
figure 1.

\begin{figure}
\begin{center}
\epsscale{1.}
\plotone{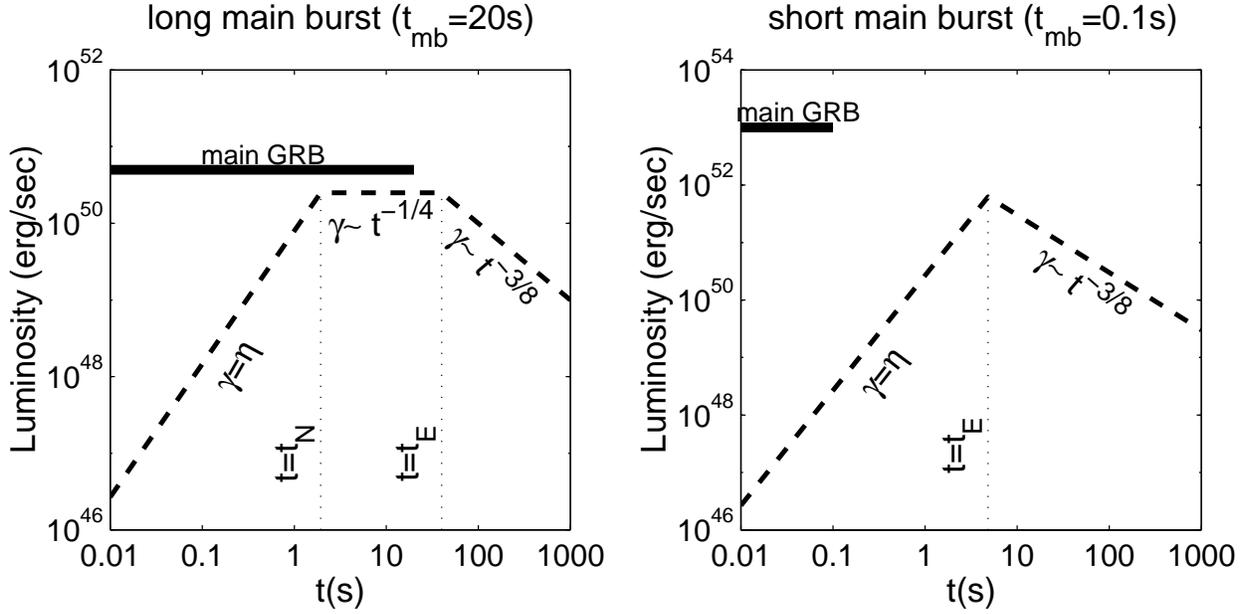}
\caption{The luminosity from the ISM as function of time in the 
Relativistic (left frame) and Newtonian (right frame) cases is drawn
in dashed line. At the early stage, the Lorentz factor is constant and
the luminosity increases due to the increase in shell area. When the
ISM has energy comparable with the total energy ($t=t_E$), a self
similar solution is established and the luminosity drops as
$t^{-1}$. If the shell is thick (typical for long main bursts) the
reverse shock becomes relativistic at $t_N$, before the self-similar
solution is established and some deceleration begins, leading to
constant luminosity. Solid line gives the luminosity of the main GRB.
Both frames use $E=10^{52}{\rm\ erg}$ and $\eta=300$. The behavior
$L \sim t^{-1}$ and $\gamma \sim t^{-3/8}$ continues up to the non-relativistic
stage which is about $30$days.
\label{fracfig}
}
\end{center}
\end{figure}

\section{Relation with the main burst}

If the main burst is produced by internal shocks, then the width of
the shell, $\Delta$, can be inferred directly from the observed main
burst duration $\Delta=c t_{mb}$. For long bursts, $t_{mb}\sim 20\sec$
and $\Delta\sim6\times 10^{11}{\rm\ cm}$ while for short bursts
$t_{mb}\sim0.1\sec$ and $\Delta\sim3\times 10^9{\rm\ cm}$. The initial
Lorentz factor must satisfy $\eta>100$ for the emission of the main
burst not to be opaque. The reverse shock is therefore likely to be
Newtonian for short bursts and might be relativistic for long
bursts. Both cases are therefore of physical interest.

If the reverse shock is relativistic, then the observed peak of the
after-glow emission is flat and overlaps the observed GRB emitted by
internal shocks. Both end after an observed time of
$\sim\Delta/c$. If the reverse shock is Newtonian, then the after-glow
peaks on $t_E$ given by Eq. (\ref{tenewtonian}) which is longer by a 
factor of 
\begin{equation}
\frac 3 2 \xi^2>1
\end{equation}
than the main burst duration $\Delta/c$. The duration and luminosity
of the main burst and the after-glow rise are shown for the Newtonian
and relativistic cases in figure 1.

In both cases, the properties of the main burst and the after-glow are
very different. The main burst is usually highly variable (depending
on the internal structure of the shell) while the after-glow, which is
due to external shocks is expected to be smooth (\markcite{SP97a}Sari
\& Piran 1997). The after-glow's spectrum should peak, in the
beginning, around $30$KeV-$10$MeV depending on the fraction of
internal energy in electrons and magnetic field (\markcite{SNP}Sari,
Narayan and Piran 1996). IF the peak energy is too high it might not
be observed in the first stage by the BATSE equipment. However later
as the ejecta decelerates the emission peak decreases in time and
should cross the soft $\gamma$-ray region.

\section{Self-Similar Solution and Properties of the Original Ejecta}

We begin with a simple consideration based upon conservation of
energy. When most of the energy has been given to the ISM, and
assuming that radiation losses are small, the energy in the shocked
ISM is constant and approximately equal to the initial kinetic energy
$E$. The shocked ISM rest mass is $M\propto R^3$. Since it was heated
by a relativistic shock its energy in the observer frame is
$\sim M\gamma^2$. Comparing this with the constant total energy of the
system $E$ we get that
\begin{equation}
\label{gammat}
\gamma\propto R^{-3/2}.
\end{equation}
This is also the scaling law for the shock wave Lorentz factor
$\Gamma$ since for a relativistic shock $\Gamma=\sqrt 2\gamma$.

The scaling law, Eq. (\ref{gammat}), implies a quantitative but
important change in the relation between $t$, $R$ and
$\gamma$. Photons that were emitted from the shock while it has
propagated a small distance $\delta R$ will be observed on time-scale
of $\delta t \sim \delta R/2\Gamma^2 c$. Integrating this over time
using the scaling law \ref{gammat} we get $t=R/8\Gamma^2c$, or
\begin{equation}
\label{rt}
t={R \over 16\gamma^2 c}.
\end{equation}
Compared with the commonly used expression $R/2\gamma^2c$
(\markcite{MeszarosRees97}M\'esz\'aros and Rees 1997,
\markcite{Waxman97a}Waxman 1997a, \markcite{WijersReesMeszaros}Wijers, Rees and
Meszaros 1997, \markcite{Waxman97b}Waxman 1997b), this expression is
factor of $8$ smaller. This difference is important when trying to fit
quantitatively the observed after-glow data. Note however that the
differential relation is independent of the deceleration, and is
therefore given by $\delta t=\delta R/4\gamma^2c$.

\markcite{BlandfordMcKee}Blandford and McKee (1976) have described an analytical solution for the 
case in which the scaling law \ref{gammat} applies. Using their solution
with several simplifications and some algebraic manipulations we get
\begin{eqnarray} 
\label{selfsimilar}
n(r,t)       &=& 4 n    \gamma   \left[ 1+16\gamma^2(1-r/R) \right]^{-5/4}     \cr
\gamma(r,t)  &=&        \gamma   \left[ 1+16\gamma^2(1-r/R) \right]^{-1/2}     \cr
e(r,t)       &=&4nm_pc^2\gamma^2 \left[ 1+16\gamma^2(1-r/R) \right]^{-17/12}
\end{eqnarray}
where $n(r,t)$, $\gamma(r,t)$ and $e(r,t)$ are, respectively, the
density, Lorentz factor and energy density of the material behind the
shock (not to be confused with the ISM density $n$ and the Lorentz
factor of material just behind the shock $\gamma(t)=\gamma(R,t)$). The
scaling laws of $R(t)$ and $\gamma(t)$ can be found using these
profiles and demanding that the total energy in the flow would be
equal to $E$:
\begin{eqnarray}
\label{scalings}
R(t)&=&\left( {17 E t \over \pi m_p n c} \right)^{1/4}
     =3.2\times 10^{16} E_{52}^{1/4}n_1^{-1/4} t_s^{1/4} {\rm\ cm} \cr
\gamma(t)&=&\frac 1 4 \left( {17 E \over \pi n m_p c^5 t^3} \right)^{1/8}
          =260 E_{52}^{1/8} n_1^{-1/8} t_s^{-3/8} 
\end{eqnarray}
This solution can serve as a starting point for detailed radiation
emission calculations and comparison with observations. The scalings
\ref{scalings} are, of course, consistent with the scalings
\ref{gammat} and \ref{rt} which were derived using conservation
of energy, but supply the exact coefficient that can not be produced
otherwise. The time for which the flow behind the shock becomes
sub-relativistic follows from Eq. (\ref{scalings}) as
\begin{equation}
t=30 E_{52}^{1/3} n_1^{-1/3} {\rm\ days}.
\end{equation}

We now turn to see the role of the particles from the original ejecta
in the flow. The number of protons (or electrons) in the ejecta is
$E/\eta m_p c^2$, and is larger than the number of ISM protons swept
by the shock ($4\pi R^3 n/3$) up to the time
\begin{equation}
t=\left( \frac {3^4}{17^3 4^4\pi}\frac{E}{nm_p\eta^4c^5} \right)^{1/3}=
850\ E_{52}^{1/3}n_1^{-1/3}\eta_{300}^{-4/3} {\rm\ s}
\end{equation}
We now use the solution \ref{selfsimilar} to determine the evolution
of the original ejecta material (or any other fluid element). Taking
the derivative of the fluid Lorentz factor along its line of motion we
get
\begin{equation}
\frac {d\gamma_e} {dt}=-\frac 7 8 \frac {\gamma_e} t
\end{equation}
so that a fluid element which had a Lorentz factor $\gamma_0$ at
$t_0$, will have
\begin{equation}
\gamma_e(t)=\gamma_0 \left( \frac t {t_0} \right)^{-7/8}.
\end{equation}
The exponent $-7/8$ shows a fast deceleration relative to the shock
deceleration exponent $-3/8$. For the Newton case, the ejecta had a
Lorentz factor $\eta$ at the beginning of the self-similar stage at
time $T_E$ given by Eq. (\ref{tenewtonian}), therefore reaching
sub-relativistic velocities after
\begin{equation}
t_{sr}=t_E \eta^{8/7}=0.9 \ E_{52}^{1/3} n_1^{-1/3} \eta_{300}^{-32/21} {\rm\ hours}
\end{equation}
For the relativistic case we get
\begin{equation}
t_{sr}=t_E \gamma^{8/7}(t_E)=2.6 \ E_{52}^{1/7} n_1^{-1/7} 
\left( \frac \Delta {6\times 10^{11}{\rm cm}}\right)^{4/7} {\rm\ hours}.
\end{equation}
This time scale of $\sim 2$hours is much shorter than the time in
which the whole flow becomes relativistic $\sim 30$days. It is
therefor expected that most of the afterglow radiation, that was seen
on time scale of days and even month, is not related to the particles
of the initial ejecta but to the shocked ISM.
\section{Radiative Corrections}

In the previous sections we have assumed that the energy in the system
is constant.  This assumption can not be strictly correct since the
radiation takes some energy from the relativistic shell. We define
$\epsilon_e$ to be the fraction of the internal energy that is
radiated, and lost from the system. Typically this should be the
fraction of energy given by the shock to electrons and is estimated to
be around $10\%$ (\markcite{Waxman97a}\markcite{Waxman97b}Waxman
1997a,b). This number seems to be negligible, and therefore the energy
loss was neglected by previous analyses. However the deceleration
occurs over several orders of magnitude in time and Lorentz factor and
the fireball energy $E$ is given again and again to newly heated
material, leading to more and more energy losses.

The energy loss rate during the deceleration is given by $4096\pi c^5
n m_p \gamma^8 t^2$ (the coefficient is different from Eq. 2, due to
the relation $R=16\gamma^2 c t$ which is relevant in the deceleration
stage) multiplied by $\epsilon_e$. Substituting the expression for
$\gamma(t)$ from the self similar solution (Eq. \ref{scalings}) we get
\begin{equation}
\frac{dE}{dt}=-\frac{17}{16}\epsilon_e\frac{E}{t},
\end{equation}
so that 
\begin{equation}
\label{et}
E(t)=E_0\left(\frac t {t_0} \right)^{-17\epsilon_e/16}.
\end{equation}
Since the observed initial time of the afterglow is about $10$s then
after about a week the energy is reduced by a factor of $\sim3$ for
$\epsilon_e=0.1$ or a factor of $\sim30$ if $\epsilon_e=0.3$. These
factors must be taken into account given the accuracy of current data.

The derivation of the above exponent, $-17\epsilon_e/16$, used the
exact coefficients in Eq. (\ref{scalings}), which were obtained from
the self-similar solution. Without this solution the exponent could
only be estimated approximately. Note that the use of
Eq. (\ref{scalings}) is valid as long as $\epsilon_e$ is small enough
that the energy loss could be approximated as a small ``radiative
correction''.

This radiation losses will also slightly effect the scaling of the
shock radius and Lorentz factor as function of time. The approximate
scaling including the radiation losses can be obtained by substituting
Eq. (\ref{et}) into Eq. (\ref{scalings}).

\section{Discussion}
We have explored the early evolution of the interaction of a
relativistic shell with the ISM. If the main GRB is short enough,
separation is expected between the main burst and the afterglow
luminosity peak, while if it is long enough an overlap is expected.
This property might be detectable in BATSE's data.

We have used the self similar solution derived by
\markcite{BlandfordMcKee}Blandford and McKee (1976) to obtain an explicit
expression for the radial profile in the self similar stage. This
solution can be used in further analyses when considering a more
detailed calculation of the radiation from the heated ISM.

A relation between the shock's radius, the material Lorentz factor and
the observed time was found to be $t=R/16\gamma^2c$ instead of the
commonly used expression $t=R/2\gamma^2c$ due to the fact that ISM is
collected so the shock moves faster than the material behind it, and
due to the deceleration of the shell, having higher Lorentz factor at
earlier time. This relation was obtained assuming that the radiation
is emitted from the shock front. On long time scales $\sim 1$day, when
the cooling of electrons is not fast enough it might be that the width
of the radiating zone will smear the observed radiation over longer
time scales.

The role of energy loss due to the radiation was found to be
non-negligible even if the part of the internal energy that is
radiated at each time is small. Thus radiation can reduce the total
energy in the system after a week by a factor of $3$ if
$\epsilon_e=0.1$ or a factor of $30$ if $\epsilon_e=0.3$.

The data of GRB970228 and GRB970508 fit radiation models, with in a
factor of 2, without taking into account radiation losses
(\markcite{Waxman97a}\markcite{Waxman97b}Waxman 1997a,b). We can
therefor roughly estimate $\epsilon_e \le 0.1$. On time scale which is
more than $\sim 1$day, the electron's cooling time is long so only a
small fraction of their energy is radiated. Since most of the
observations where made after $\sim 1$day, the fraction of the energy
that is given to the electrons can be high (more than $10\%$) without
leading to considerable energy loss, and with no effect on the
observations made after $\sim 1$day. However, in such a case, the
energy losses at earlier time will be considerable and will therefor
require a much higher initial energy.

\acknowledgments

The author thanks The Institute for Advanced Studies for warm
hospitality and Eli Waxman, Pawan Kumar, John Bahcall, Tsvi Piran,
Jonathan Katz, and Shiho Kobayashi for helpful discussions.

\end{document}